\documentstyle[12pt]{article}

\def\tstrut{\vrule height 1em depth 0.5em width 0pt}
\newcommand{\s}{$\bar{s}s$}
\newcommand{\p}{$\bar{p}p$}
\newcommand{\f}{$\phi$}
\newcommand{\an}{annihilation}
\newcommand{\na}{na\"{\i}ve}
\newcommand{\ap}{antiproton}
\newcommand{\om}{$\omega$}
\newcommand{\ten}{$f'_2 (1525) $}

\newcommand{\beq}{\begin{equation}}
\newcommand{\eeq}{\end{equation}}
\def\eqref#1{(\ref{#1})}
\def\naive{na\"{\i}ve}

\begin{document}
\begin{flushright}
{hep-ph/9909235} \\
{CERN-TH/99-236} \\
TAUP--2585-99
\end{flushright}
\vskip1.5cm
\begin{center}
{\Large\bf Hadronic Probes of the Polarized Intrinsic\tstrut} \\
{\Large\bf Strangeness of the Nucleon}
\end{center}
\medskip
\begin{center}
{\large J. Ellis$^{a}$, M. Karliner$^{b}$,
D.E. Kharzeev$^{c}$
  and
M.G. Sapozhnikov$^{d}$}
\end{center}
\vskip1cm
\begin{center}
$^{a)}$ Theory Division, CERN, Geneva, Switzerland\\
\vskip0.3cm
$^{b)}$ School of Physics and Astronomy,\\
Raymond and Beverly Sackler Faculty of Exact Sciences\\
Tel Aviv University, Tel Aviv, Israel\\
\vskip0.3cm
$^{c)}$ RIKEN BNL Research Center, \\
Brookhaven National Laboratory,\\
 Upton, New York, USA\\
\vskip0.3cm
$^{d)}$ Joint Institute for Nuclear Research, Dubna, Russia
\end{center}
\vskip0.5cm
\begin{abstract}
We have previously interpreted the various
large apparent violations of the \na\ Okubo-Zweig-Iizuka
(OZI) rule found in
 many channels in $\bar{p}p$ \an\ at LEAR as evidence for
an intrinsic polarized \s\
component of the nucleon wave function. The model is further
supported by new data from LEAR and elsewhere. Here we discuss
in more detail the possible form of the \s\
component of the nucleon wave function, interpret the new data
and clarify the
relative roles of strangeness shake-out and rearrangement,
discuss whether alternative interpretations are still allowed
by the new data, and propose more tests of the model.
\end{abstract}
%\vskip1.4cm
\vfill
%CERN--TH.7326/94\\
%TAUP--2177/94\\
%December 1994
\vfill\eject

\section{Introduction}

    The possible strange quark content of the nucleon is currently
of considerable experimental and theoretical interest. It should not
surprise anyone that the nucleon wave function might contain a
substantial \s\ component. The analysis of QCD sum rules
shows that the condensate of \s\ pairs
in the vacuum is not small, but is comparable with the condensate
of the light quarks~\cite{Iof.81}.
The presence
of \s~ pairs in the nucleon was first indicated by
measurements of the $\pi$-nucleon $\sigma$ term~\cite{Gasser:1991ce}, and by
charm production in deep-inelastic neutrino
scattering~\cite{Adams:1999sy}.
Over the past decade,
the EMC and successor experiments with polarized lepton beams and
nucleon targets (SMC, E142, E143, E154, HERMES)~\cite{Duren:1999vs}
 have
not only confirmed that such \s\ pairs
are present, but also indicated that they are polarized. This
latter observation has stimulated great theoretical interest, with
chiral soliton models~\cite{Brodsky:1988ip}
and ideas based on the
axial
$U(1)$
anomaly~\cite{Efremov:1988zh,Altarelli:1988nr,Carlitz:1988ab}
providing competing focuses for the debate~\cite{Vogelsang:1999up}.

    Within this general phenomenological ferment, attention has
been drawn back to the Okubo-Zweig-Iizuka (OZI) rule~\cite{OZI} and its
applicability to processes involving baryons~\cite{Ell.89}. The OZI rule
postulates that diagrams with disconnected quark loops, e.g., for
$\phi$ decay into $\pi \pi \pi$ or
$J/\psi$ decay into light hadrons, should be suppressed compared
to connected quark diagrams, e.g., for $\rho$ decay into $\pi \pi$.
Of particular interest has been the OZI-inspired suggestion that,
in reactions involving pions and nucleons,
mesons with a predominant \s\ content, such as
the $\phi$ and the $f'(1525)$, should be produced only via their
$\bar u u + \bar d d$ contents, i.e., via the departures from
ideal mixing in their wave functions. However, this prediction would hold
only if the pion and nucleon wave functions contained negligible
fractions of \s\ pairs, which, in the case of nucleons, is not supported
by
the experiments mentioned in the first paragraph.

    Indeed, as was pointed out in~\cite{Ell.89}, early data
on $\phi$ production indicated excesses above predictions
based on the departure from ideal mixing in the $\phi$ wave
function, especially in low-energy \p\ annihilation.
The interpretation proposed in~\cite{Ell.89} was that
there was additional production by diagrams that are allowed
by the OZI rule if the nucleon wave function contains \s\
pairs. Such a model would predict channel-dependent
departures from predictions based on ideal mixing.
Other interpretations of the \p\ data have
included the existence of a four-quark
$\bar q q \bar s s$ meson $C$ in one particular partial
wave~\cite{Dov.89},
and the importance of rescattering via $K \bar K$ and $K \bar K^*
+ \bar K K^*$ intermediate states~\cite{Loc.94,Lev.94}.

A wealth of new high-statistics data in various \p\
annihilation channels have recently become available, coming
principally from the LEAR experiments OBELIX, the
Crystal Barrel and JETSET. These provide information on several
final states including $\phi \gamma$, $\phi \pi$, $\phi \eta$,
$\phi \pi \pi$,
$f' \pi$ and $\phi \phi$, in
different experimental conditions which allow
initial-state spin and orbital angular momentum states to be
distinguished. These data provide unambiguous evidence for the
failure of the
application of the OZI rule with a \na\ nucleon wave function,
and provide many hints for the formulation of a more satisfactory
model.

    In a previous paper~\cite{Ell.95}, we proposed a model
based on a nucleon wave function containing negatively polarized
$\bar s s$ pairs, as suggested by the EMC and successor
experiments. This model opened up the possibilities of $\phi$
production via shake-out and rearrangement diagrams, whose
importance would depend on the initial nucleon-antinucleon
states. Some aspects of the model were not specified at first:
for example, the partial waves of the $\bar s s$ pair relative
to the core of the nucleon wave function, and between the $s$
and $\bar s$, were left open. Nevertheless, this model
explained satisfactorily many subtle qualitative aspects of $\phi$
production in $\bar p p$ annihilation, as well as making
interesting predictions for $\phi$ production elsewhere and
for other experiments. In particular, the model of~\cite{Ell.95}
was extended to the reaction $\bar p p \rightarrow
\Lambda \bar \Lambda$ in~\cite{Alb.95}, where arguments were given on the
basis of
chiral symmetry that the $\bar s s$ pair in the nucleon wave
function might be in a $^3$$P_0$ state. Also,
the model of~\cite{Ell.95} was applied in \cite{Ek.95}
to make predictions for
$\Lambda$ longitudinal polarization in the target fragmentation
region for deep-inelastic lepton scattering.

    Even more new data from LEAR and elsewhere have appeared
since \cite{Ell.95}, and they seem to bear out the general
features of the model proposed there. The purposes of this
paper are the following:
\begin{itemize}
\item
To refine the arguments of \cite{Ell.95} and \cite{Alb.95} on
the possible form of the $\bar s s$ component in the nucleon
wave function;
\item
To interpret the new data in the light of this model, clarifying
the relative roles of shake-out and rearrangement;
\item
To discuss whether other interpretations are still tenable, in the
light of the more detailed experimental information now available;
and
\item
To propose more tests of the model.
\end{itemize}

\section{The Na\"{\i}ve OZI Rule and its Apparent \\Violation}

There are different formulations of the OZI rule~\cite{OZI}
(for a review, see \cite{Lipkin}), which is stated as the suppression of
reactions with disconnected quark lines. In order to make quantitative
statements it is convenient to use the formulation of Okubo~\cite{Oku.77}.
Consider the production of
$\bar{q} q$ states in the interactions of hadrons
\begin{eqnarray}
A+B\longrightarrow C + q\bar{q} ~~~~\mbox{for}~~q=u,d,s
\end{eqnarray}
where the hadrons $A,B$ and $C$ consist of only light quarks.
The na\"{\i}ve OZI rule demands that
\begin{eqnarray}
Z = \frac{\sqrt{2} M(A+B\rightarrow C+s\bar s)}
{M(A+B\rightarrow C+u\bar u) + M(A+B\rightarrow C+d\bar d)}=0 \label{OZI}
\end{eqnarray}
where $M(A+B\rightarrow C+q\bar{q})$ are the amplitudes of the
corresponding processes.

This means that, if the $\phi$ meson were a pure $\bar{s}s$ state,
it could not be produced in the interaction of ordinary
light-quark hadrons.
The OZI rule in Okubo's form~\cite{Oku.77} strictly forbids the production
of
strangeonia or charmonia from hadrons composed only of $u$ and $d$ quarks.
The only way to form a $\phi$, according to this rule, is through the
admixture
of the light quarks in the $\phi$ wave function. This admixture is
parametrised
by the parameter $\delta \equiv \Theta - \Theta_i$,
where $\Theta$ and $\Theta_i$ are the physical and ideal mixing angles,
respectively. If the
mixing angle is ideal: $\Theta_i=35.3^0$, $\phi$ is a pure $\bar s s$
state. In practice, the physical value of the mixing
angle $\Theta$ may be determined phenomenologically
from the masses of the mesons in the
corresponding meson nonet.

The na\"{\i}ve OZI rule of~(\ref{OZI}) may be re-written in terms
of the mixing angle $\Theta$:
\begin{eqnarray}
\frac{M(A+B\rightarrow C+\phi)}{ M(A+B\rightarrow C+\omega)} =
- \frac{ Z+ \tan(\Theta-\Theta_i)}{1-Z\tan(\Theta-\Theta_i)} \label{MAB}
\end{eqnarray}

 If the na\"{\i}ve OZI rule is correct, i.e., $Z=0$,
then
\begin{equation}
R = \frac{\sigma(A+B \rightarrow\phi X)}{\sigma(A+B\rightarrow\omega X)}
= \tan^2{(\Theta - \Theta_i)}\label{R}
\end{equation}
which predicts a universal suppression factor for
$\phi$ production relative to $\omega$ production. However,
it is necessary to make a phase-space correction before comparing this
ratio with the experimental data.

Using the quadratic Gell-Mann-Okubo
mass formula to determine the physical
vector-meson mixing angle, one may obtain
\begin{equation}
R(\phi/\omega) = 4.2\cdot10^{-3}
\label{eq:RV}
\end{equation}
and the corresponding analysis for
tensor mesons yields
\begin{equation}
R(f_2'(1525)/f_2(1270)) = 16\cdot10^{-3}
\label{eq:RT}
\end{equation}
Alternatively, the tensor-meson mixing angle may be
determined from the \ten~ decay widths into the OZI-forbidden
$\pi\pi$ and OZI-allowed $\bar{K}K$ modes:
\begin{equation}
R=\frac{W(f_2'(1525)\to \pi\pi)}{W(f_2'(1525) \to \bar{K}K)} =
(2.6\pm0.5)\cdot10^{-3}
\label{eq:RTdecay}
\end{equation}
after making the appropriate phase-space correction.

The predictions (\ref{eq:RV}) and  (\ref{eq:RT}) were tested
many times in experiments using different hadron beams.
The analysis~\cite{Nom.99} of the experiments collected in the
Durham reactions data base  shows that in $\pi N$
interactions
the weighted average ratio of cross sections of \f~ and \om~production
at different energies is
\begin{equation}
\bar{R}=\frac{\sigma(\pi N\to \phi X)}{\sigma(\pi N\to \omega X)}
= (3.3\pm0.3)\cdot10^{-3}
\label{eq:RV1}
\end{equation}
without attempting to make a phase-space correction.
Therefore, in $\pi N$ interactions, the
agreement with the na\"{\i}ve OZI rule prediction (\ref{eq:RV}) is
very good, with the result (\ref{eq:RV1})
corresponding to a value $ Z_{\pi N} = 0.9\pm0.3 \%$
for the OZI violation parameter, possibly becoming
smaller if phase space is taken into account.

The weighted average ratio of cross sections of \f~ and \om~production
at different energies
in nucleon-nucleon interaction is somewhat higher, but
still qualitatively similar to the OZI value (\ref{eq:RV}):
\begin{equation}
\bar{R}=\frac{\sigma(NN \to \phi X)}{\sigma(NN\to \omega X)}
= (14.7\pm1.5)\cdot10^{-3}
\label{eq:RV2}
\end{equation}
This corresponds to a value $ Z_{NN} = 8.2\pm0.7 \%$
for the OZI violation parameter, without attempting to make
any correction for the phase-space differences of
the final states, which would be non-universal. However, these
would always tend to increase (\ref{eq:RV2}) and hence $Z_{NN}$.
The corresponding values for
\ap~\an~ in flight are qualitatively similar:
\begin{equation}
\bar{R}=\frac{\sigma(\bar{p}p \to \phi X)}{\sigma(\bar{p}p\to \omega X)}
 =  (11.3\pm1.4)\cdot10^{-3}
\label{flight}
\end{equation}
yielding the value $Z_{\bar{p}p} = 5.0\pm0.6 \% $.

These experiments indicate that
the na\"{\i}ve OZI rule for vector meson production is
generally valid within 10$\%$
accuracy. This is not so bad for a heuristic model, bearing in mind that
the OZI prediction is based only
on the value of the mixing angle derived from meson masses, and
applies at different energies from 100 MeV till 100 GeV.
Experimental data on tensor meson production are more
scarce, but in general they also confirm the
na\"{\i}ve OZI rule prediction (\ref{eq:RT}).
\vskip0.3cm

In view of this seemingly comfortable situation,
it was a surprise when experiments
at LEAR with stopped \ap~ showed large violations of the
na\"{\i}ve OZI rule (for reviews, see~\cite{Sap.95,Ams.98}).
The largest violation observed is for the $\bar{p} p \to \phi\gamma$
channel, where the Crystal Barrel collaboration has found~\cite{Ams.98} for
the ratio of yields for the reactions
$\bar{p} + p \to \phi(\omega) + \gamma $:
\begin{equation}
R_{\gamma} \equiv {\sigma (\bar p p \rightarrow \phi \gamma )
\over \sigma ( \bar p p \rightarrow \omega \gamma)}
= (294\pm 97)\cdot 10^{-3},
\label{Rgamma}
\end{equation}
which is about 70 times larger than the
OZI prediction $R(\phi/\omega)=4.2\cdot10^{-3} $!
Another very
large apparent violation of the OZI rule was found for the
$ \bar{p} + p \to \phi(\omega) + \pi^0 $ channels,  where
the analogous quantity is
\begin{equation}
R_\pi= (106\pm 12)\cdot 10^{-3}
\label{Rpi}
\end{equation}
for annihilation in a liquid-hydrogen target~\cite{Ams.98},
and
\begin{equation}
R_\pi = (114\pm 24)\cdot 10^{-3}
\label{Rpitwo}
\end{equation}
for annihilation in a hydrogen-gas target~\cite{Abl.95b}.
These ratios are
about a factor of 30 higher than the \na~ OZI rule prediction!

One of the most striking features of the violation of the \na~
OZI rule found in the experiments at LEAR is its
strong dependence on the quantum numbers of the initial state.
This interesting effect was initially observed by the ASTERIX
collaboration \cite{Rei.91}
in the $\bar{p}p \to \phi\pi^0$ channel.
This reaction is allowed from two $\bar{p}p$
initial states, namely $^3S_1$ and $^1P_1$~, but $\phi$ mesons were
observed in the sample with the dominant S-wave content
and were not seen at all in the sample corresponding to P-wave
annihilation:
\begin{eqnarray}
 Br(\bar{p}p\to \phi\pi^0, ~^3S_1) & = & (4.0\pm0.8)\cdot 10^{-4}~,
\\
 Br(\bar{p}p\to \phi\pi^0, ~^1P_1) & < & 0.3\cdot 10^{-4}
\label{channel}
\end{eqnarray}
which is a challenge for model interpretations.

To illustrate that the amount of apparent OZI-rule violation
observed in \p~ \an~at rest is quite unusual,
we compare the $\phi/\omega$
ratios observed in \p ~\an~ with
corresponding channels in $J/\psi$ decays:
\begin{equation}
R_{J/\psi} \equiv \frac{B(J/\psi \to \phi \pi^+\pi^-)}{B(J/\psi \to \omega
\pi^+\pi^-)} =
(111 \pm 23) \cdot10^{-3}
\label{Jpsi}
\end{equation}
Since the initial state is light-flavour-neutral, this gives the ballpark
of the
`natural' $\phi/\omega$ ratio in a case where both $\phi$ and $\omega$ are
produced via disconnected quark-line diagrams. It therefore seems
surprising that in  \p~\an~ at rest the $\phi/\omega$ ratio is of
the same order of magnitude as in $J/\psi$ decays,  despite
the additional possibility of the $\omega$ being produced by the
rearrangement
of light quarks in the $\bar p p$ annihilation case.

To add to the puzzle,
there are channels for $\phi$ meson production
in $\bar{p}p$ annihilation at rest where
no OZI-rule violation was observed. For instance,
it was found for annihilation in liquid hydrogen that
$R(\phi \eta/\omega \eta) = (4.6\pm 1.3)\cdot 10^{-3}$~\cite{Ams.98},
compatible with the \na\ OZI rule. Moreover,
no large enhancement of \f~ production was observed in the ratio
$R(\phi \omega/\omega \omega) = (19\pm 7)\cdot 10^{-3}$ ~\cite{Sap.95},
and $R(\phi \rho/\omega \rho) = (6.3\pm 1.6)\cdot 10^{-3}$~\cite{Sap.95}
is also compatible with the \na\ OZI rule.

One possible interpretation of these facts was proposed in~\cite{Ell.95},
where it was
assumed that the nucleon wave function contains polarized $\bar s s$
pairs. Then, there are additional classes of connected
quark-line diagrams, and the
observed OZI violation is only apparent. It was
further proposed that the strong dependence on the
initial-state quantum numbers is due to polarization of the strange sea.
We next turn to a more detailed study of this proposal.

\section{The $\bar{s}s$ Component of the Nucleon Wave Function}

There are many different possibilities for the quantum
numbers of the \s~ component in the nucleon wave function~\cite{Iof.90}.
We attempt a simplified description of the proton
as a combination of $uud$ and \s~ clusters and assume, for simplicity,
that the quantum numbers of the
$uud$ cluster are the same as for the proton, namely $J^P = 1/2^+$.
One may then still explore an infinite spectrum of possible
quantum numbers of the \s~ quark cluster. The possibilities
involving the lowest relative partial waves
are those shown in Table 1.

\begin{table}[hbt]
\label{tab:jpc}
%\begin{tabular}{rrrrr}
\begin{tabular*}{\textwidth}{@{}l@{\extracolsep{\fill}}rrrrr}
%\begin{tabular*}{\textwidth}{@{}l@{\extracolsep{\fill}}rrrrr}
\hline
S & L & j & $J^{PC}$ & State\\
\hline
0 & 0 & 1 & $0^{-+}$ &$^1S_0$ `$\eta$'\\
1 & 0 & 1 & $1^{--}$ &$^3S_1$ `$\phi$'\\
1 & 1 & 0 & $0^{++}$ &$^3P_0$ \\
1 & 1 & 0 & $1^{++}$ &$^3P_1$ \\
0 & 1 & 0 & $1^{+-}$ &$^1P_1$ \\
\hline
\end{tabular*}
%\end{tabular}
%\end{table*}
\vspace*{0.2cm}
\caption{{\it Possible
quantum numbers of the \s~ cluster in the nucleon.
We denote by $\vec{S}$ and $\vec{L}$ the
total spin and orbital angular momentum of the \s~pair,
$\vec{J} \equiv \vec{L}+\vec{S}$, and
the relative angular momentum between
the \s~ and $uud$ clusters is $\vec{j}$.}}
\end{table}

We see from Table 1 that the \s\ could be stored in the nucleon with the
quantum numbers of either the $\eta$ or the $\phi$
if the relative angular momentum
between the \s~ and $uud$ clusters is $j=1$. However, if
$j=0$, the quantum numbers of
\s~ pair may be
different, and the vacuum quantum
numbers $J^{PC}=0^{++}$ become an option. It is clear that
the predictions of the model will depend drastically
on the assumption about the \s~ quantum numbers.

The assumption that the \s~ pair has the
quantum numbers of $\phi$ meson may seem attractive
{\it a priori}, but leads to serious problems.
 In this case one would expect additional
\f~ production due to {\em shake-out} of the hidden strangeness
\cite{Ell.95} stored in
the nucleon, but
it is not clear how to explain the strong dependence of the
$\phi$ yield on the joint quantum numbers of the pair of
annihilating nucleons,
as first observed by the ASTERIX Collaboration~\cite{Rei.91}.
Moreover, the shake-out of the $\phi$ mesons stored in the nucleon
would lead to an apparent violation  of the
OZI rule in {\it all} reactions.
The fact that apparent OZI violation is not seen in
some reactions, and is relatively small in some others, as reviewed
in the previous Section,
seems to exclude this possibility. It seems difficult to
explain in this picture why
the OZI rule should be obeyed within 10 \% accuracy in some
cases, but violated by factors of the order of 30 or more in other cases.

Similar arguments were used in~\cite{Dov.90}, where
it was demonstrated
that the experimental
data on the production of $\eta$ and $\eta'$ mesons exclude $0^{-+}$
quantum numbers for the \s~ admixture in the nucleon wave function.

As an alternative, it was argued in~\cite{Alb.95}
that the $\bar s s$ cluster in the nucleon
might be negatively polarized, due to the interaction of the light
valence quarks with the QCD vacuum. Due to their chiral dynamics,
the interaction between quarks and antiquarks is strongest in
the pseudoscalar $J^{PC}=0^{-+}$ sector. This strong attraction
in the spin--singlet pseudoscalar channel between a light valence
quark in the proton wave function and a strange antiquark from
the QCD vacuum could result in the spin of the strange antiquark
being aligned antiparallel to the spin of the light quark, and
hence, on average, to the proton spin. On the other hand, the
QCD vacuum is known to contain quark-antiquark pairs with the
quantum numbers $J^{PC}=0^{++}$, i.e., in a $^3P_0$ state.
A QCD sum-rule analysis~\cite{Iof.81} indicates that
the condensate of the strange
quarks in the vacuum is not small compared with the condensate
of the light quarks:
\begin{equation}
< 0\left|\bar{s}s \right|0> = (0.8\pm0.1)< 0\left|\bar{q}q \right|0>,~~
q=(u,d)
\end{equation}
If the $\bar s s$ pair in the nucleon can be regarded as a
$^3P_0$ vacuum excitation, the spin of the $s$
quark should be aligned with that of the $\bar s$ antiquark,
so to preserve the vacuum
quantum numbers $J^{PC}=0^{++}$, and hence should also be aligned opposite
to the nucleon spin.
It is important to stress that the $\bar s s$ pair with $^3P_0$
quantum numbers  is not itself polarized, being a scalar. Rather,
a chiral non--perturbative interaction selects
one projection of the  total spin of the $\bar s s$ pair on the direction
of the nucleon spin, namely that with $S_z = -1$~\footnote{We comment,
in this connection,
that QCD instantons have been shown to
induce just such a {\em negative} quark
polarization~\cite{FS.91,Koc.98}.}.
Since the density of $\bar{s}s$ pairs in the QCD vacuum is quite high,
one might expect that the effects of the polarized strange quarks in
the nucleon could also be non-negligible, as suggested by
the polarized deep-inelastic scattering data: see also the discussion in
\cite{Iof.90}~\footnote{These strange
quarks should not be considered to be constituent quarks
in some five-quark configuration of the nucleon. Rather, they
should be considered as components of a constituent quark.}.

The above discussion of the sea quantum numbers has an immediate 
application to the question of contributions of nucleon constituents to 
its total spin.
Consider the sum rule for the total helicity of the nucleon,
\beq
{1\over2} = {1\over2} \sum_q \Delta q + L_z + \Delta G 
\label{angmom_sumrule}
\eeq
where $\Delta q$, $L_z$ and $\Delta G$ are respectively
the quark helicity, the orbital
angular momentum of all partons, and the gluon helicity.
It is useful to rewrite \eqref{angmom_sumrule}, separating the sea 
and valence quark contributions to $\Delta q$ and $L_z$
\beq
{1\over2} = {1\over2} \sum_q \Delta q^{val} + \Delta q^{sea}
+L_z^{val} + L_z^{sea} + L_z^{glue} + \Delta G
\label{angmom_sumrule_a}
\eeq
If the sea quarks in the nucleon are in $^3P_0$ state, then
the total helicity they carry
is equal to their orbital angular momentum,
$\Delta q^{sea} = -L_z^{sea}$.
The sum rule then becomes
\beq
{1\over2} = {1\over2} \sum_q \Delta q^{val} 
+L_z^{val} + L_z^{glue} + \Delta G
\label{angmom_sumrule_b}
\eeq
and since it is likely that $L_z^{val}=0$, we have
\beq
{1\over2} = {1\over2} \sum_q \Delta q^{val} + L_z^{glue} + \Delta G
\label{angmom_sumrule_c}
\eeq
The leading-order evolution equations tell us that 
\beq
{d \Delta G \over d\log Q^2} = - {d  L_z^{glue} \over d\log Q^2} 
\label{glue_evolution}
\eeq
i.e., to this order the evolution preserves the initial $(L_z^{glue}
+\Delta G)$ one starts with at $Q^2=Q^2_0$. If one starts with 
a ``bare" nucleon in the lowest Fock state $|uud\rangle$, with all the
helicity carried by its three valence quarks, then 
$L_z^{glue} = -\Delta G$. Plugging this into
\eqref{angmom_sumrule_c}, we recover the \naive\ quark model result 
\beq
{1\over2} = {1\over2} \sum_q \Delta q^{val}
\label{angmom_sumrule_d}
\eeq
Thus we see explicitly how the helicity sum rule is satisfied. 
As the bare nucleon gets ``dressed" through the addition of
sea quarks and gluons, each such addition leaves the total
helicity unchanged, since the respective helicity and orbital angular 
momentum contributions cancel each other.

\section{Production of $\bar{s}s$ Quarkonium Systems}

We now consider the production of \s\ strangeonia in
$NN$ or $\bar{N}N$ interactions,
assuming the picture outlined in the previous Section,
namely that the nucleon wave function
contains an admixture of negatively
polarized $\bar{s}s$ pairs in a $^3P_0$ state, i.e., $J^{PC}=0^{++}$.

The shake-out of such pairs will not create a \f\
or tensor \ten\ meson, but rather
a scalar strangeonium state. No concrete candidate for this state is
firmly established (see~\cite{PDG} for a discussion), though
the lightest $\bar{s}s$ scalar may have a mass around 1700 MeV.
In this case, the shake-out of the scalar \s\ pair from the nucleon
will be
a source of channels with open strangeness, such as $\bar{K}K$ and
$KK^*$. Other, identifiable
\s\ systems should be produced by processes in which
strange quarks from {\it both} nucleons participate, and therefore
depend on the quantum numbers of the initial $ N N$ state.

\vspace{1cm}
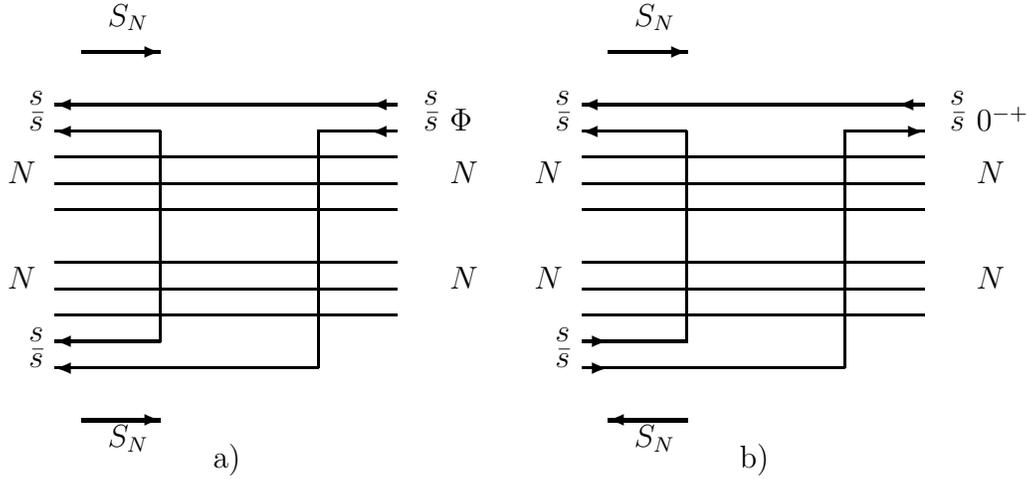
\begin{figure}[htb]
%\begin{minipage}[t]{180mm}
%\framebox[179mm]{\rule[-26mm]{0mm}{52mm}}
% Picture of rearrangement ss in NN, triplet
%\makebox{90,80}[l]{
\setlength {\unitlength} {0.7mm} \thicklines
\begin{picture}(180,70)(0,10)
%proton 1, low
\put(60,20){\vector(-1,0){50}}
\put(30,25){\vector(-1,0){20}}
\put(10,30){\line(1,0){65}}
\put(10,35){\line(1,0){65}}
\put(10,40){\line(1,0){65}}
%proton 2, upper
\put(70,70){\vector(-1,0){60}}
\put(30,65){\vector(-1,0){20}}
\put(10,60){\line(1,0){65}}
\put(10,55){\line(1,0){65}}
\put(10,50){\line(1,0){65}}
%annihilation, vertical lines
\put(30,65){\line(0,-1){40}}
\put(60,20){\line(0,1){45}}
%final lines
\put(75,70){\vector(-1,0){5}}
\put(75,65){\vector(-1,0){5}}
\put(60,65){\line(1,0){10}}
\put(75,35){\line(-1,0){20}}
\put(75,30){\line(-1,0){20}}
%texts
\put(40,1){a)}
\put(5,20){$\bar{s}$}
\put(5,25){$s$}
\put(5,65){$\bar{s}$}
\put(5,70){$s$}
\put(80,65){$\bar{s}$}
\put(80,70){$s$}
\put(85,65){$\Phi$}
\put(1,35){$N$}
\put(1,55){$N$}
\put(85,35){$N$}
\put(85,55){$N$}
\put(20,85){$S_{N}$}
\put(20,5){$S_{N}$}
%N1 and N2 spins
\linethickness{0.5mm}
\put(15,10){\vector(1,0){15}}
\put(15,80){\vector(1,0){15}}
%\end{picture}

% Picture of rearrangement ss in NN, singlet

\setlength {\unitlength} {0.7mm} \thicklines
%\begin{picture}(80,70)(0,10)
%proton 1, low
\put(110,20){\vector(1,0){5}}
\put(160,20){\line(-1,0){45}}

\put(110,25){\vector(1,0){5}}
\put(130,25){\line(-1,0){15}}

\put(110,30){\line(1,0){65}}
\put(110,35){\line(1,0){65}}
\put(110,40){\line(1,0){65}}
%proton 2, upper
\put(170,70){\vector(-1,0){60}}
\put(130,65){\vector(-1,0){20}}
\put(110,60){\line(1,0){65}}
\put(110,55){\line(1,0){65}}
\put(110,50){\line(1,0){65}}
%annihilation, vertical lines
\put(130,65){\line(0,-1){40}}
\put(160,20){\line(0,1){45}}
%final lines
\put(175,70){\vector(-1,0){5}}
\put(170,65){\vector(1,0){5}}
\put(160,65){\line(1,0){10}}
\put(175,35){\line(-1,0){20}}
\put(175,30){\line(-1,0){20}}
%texts
\put(140,1){b)}
\put(105,20){$\bar{s}$}
\put(105,25){$s$}
\put(105,65){$\bar{s}$}
\put(105,70){$s$}
\put(180,65){$\bar{s}$}
\put(180,70){$s$}
\put(185,65){$0^{-+}$}
\put(101,35){$N$}
\put(101,55){$N$}
\put(185,35){$N$}
\put(185,55){$N$}
\put(120,85){$S_{N}$}
\put(120,5){$S_{N}$}
%N1 and N2 spins
\linethickness{0.5mm}
\put(130,10){\vector(-1,0){15}}
\put(115,80){\vector(1,0){15}}
\end{picture}
\vspace*{1cm}
\caption{\it Production of the \s~ mesons in $NN$ interaction
from the spin-triplet (a) and spin-singlet (b) states.
The arrows
show the direction of spins of the nucleons and strange quarks.}
\label{fig:8}
\end{figure}

One example of
such a {\em rearrangement} diagram is shown in Fig.~\ref{fig:8}.
If the nucleon spins are parallel, as in Fig.~\ref{fig:8}a,
then, in the proposed model, the spins of
the $\bar{s}$ and $s$ quarks in both nucleons are also parallel.
If the polarization of the strange quarks is not changed during the
interaction, then the $\bar{s}$ and $s$ quarks will have parallel
spins in the final state. The total spin of the \s\ quarks will
therefore be $S=1$. If their relative orbital momentum is $L=0$,
the produced
strangeonium will have the \f\ quantum numbers.
On the other hand, if $L=1$, the produced strangeonium will
have the $f'_2(1525)$ quantum numbers.

In the case where the initial $ NN$ state is a spin-singlet, the
spins of the \s\ pair
in different nucleons are antiparallel, and
rearrangement diagrams like that in Fig.~\ref{fig:8}b may produce
a final-state \s\ system with total spin $S=0$. This
means
that, for $L=0$, a strangeonium state with the pseudoscalar quantum
numbers $0^{-+}$ is produced.

Therefore, in our picture there are always diagrams corresponding to the
rearrangement of the \s\ pairs between different nucleons, but this
does not mean that the \s\ pair is created as a $\phi$ meson.
The predictions of the polarized-strangeness model
are quite definite:
\begin{itemize}
\item The \f\ should be produced more strongly from the $^3S_1$ state,
\item The \ten\  should be produced more strongly from the $^3P_J$ states,
\item Spin-singlet initial states favour the formation of
pseudoscalar strangeonia.
\end{itemize}
This mechanism could therefore explain why the $\phi/\omega$ ratio is not
universal in different initial-state
channels for $\bar N N$ annihilation at rest.
Note also that these rules should hold
for nucleon-nucleon scattering, as well
as for antiproton-proton annihilation

\section{Comparison with the Experimental Data}

As we now discuss, new experimental data on $pp$, $pn$ and $\bar{p}p$
interactions provide valuable information that tends to verify many
predictions of the polarized strangeness model.\\
~\\
\noindent
$\bullet$ {\bf Enhancement of \f\ production from spin--triplet
states}\\
~\\
\noindent
The OBELIX collaboration has recently measured
the $\bar{p}p \to K^+K^-\pi^0$
channel for annihilation of stopped antiprotons in liquid,
and in gas at NTP and at 5 mbar pressure~\cite{Pra.98}.
These data enabled the branching ratios of the
$\bar{p}p \to \phi\pi^0$ channel to be
determined for definite initial states:
\begin{eqnarray}
 Br(\bar{p}p\to \phi\pi^0, ~^3S_1) & = & (7.57\pm0.62)\cdot 10^{-4}~,
\label{3s}\\
 Br(\bar{p}p\to \phi\pi^0, ~^1P_1) & < & 0.5\cdot 10^{-4}
\label{1p}
\end{eqnarray}
where the latter is a 95 \% confidence-level upper limit.
The indication  of a
strong dependence of the $\phi\pi^0$ production rate on
the quantum numbers of the initial $\bar{p}p$ state,
obtained earlier by the ASTERIX collaboration~\cite{Rei.91},
has therefore been confirmed with statistics higher by
a factor of 100.
The branching ratio of the $\phi\pi^0$ channel from
the $^3S_1$ initial state is at least 15 times larger than that
from the $^1P_1$ state.
The polarized nucleon strangeness explains this
remarkable selection rule, a feature not shared by other
theoretical models.

It is important to note that
the observed tendency exists only for \f\ meson production,
whereas P-wave \om\ meson production is quite plentiful, as was observed
in measurements of the antineutron-proton
annihilation~\cite{Fil.98}. The cross section of the
$\bar{n}p \to \phi (\omega) \pi^+$ channel was measured
for antineutron momenta in the range 50-405 MeV/c. It turns out that the
$\phi \pi^+$
cross section drops with energy
in the same way as the S-wave. On the other hand, the
$\omega \pi^+$
cross section  does not decrease so rapidly with energy.
Moreover, a Dalitz-plot fit for the $\omega \pi^+$ final state
demands a significant P-wave contribution.
The branching ratio of
$\omega \pi^+$ channel from the $^3S_1$ final state is
$B.R.(^3S_1)=(8.51\pm0.26\pm0.68)\cdot10^{-4}$, whereas
that from the
$^1P_1$ final state is only three times less:
$B.R.(^1P_1)=(3.11\pm0.10\pm0.25)\cdot10^{-4}$.
This tendency is in sharp contrast with the same branching ratios for
the $\phi\pi$ final state (\ref{3s})-(\ref{1p}).
It was also found that the ratio
$R = Y(\phi\pi^+)/Y(\omega \pi^+)$  decreases with increasing
antineutron energy, as predicted by the polarized-strangeness
model \cite{Ell.95}.
In this case, the energy dependence of $\phi$ production should follow
the energy dependence of the admixture of the $^3S_1$ initial state, which
declines from the
$\bar{n}p$ threshold. On the other hand, in the case of $\omega$
production, both S- and P-wave amplitudes are possible and
the P-wave admixture rapidly increases from the threshold, so we
expect $R = Y(\phi\pi^+)/Y(\omega \pi^+)$ to decrease with increasing
antineutron energy.

Analogous differences in the
angular distributions of \f\ and \om\
mesons produced in $pp \to pp \phi(\omega)$ interactions
were observed by the DISTO collaboration~\cite{Bal.98}.
Whereas the \f\ meson was found to be formed predominantly in an S-wave
state relative
to the $pp$ system, the \om\ angular distribution reveals
contributions from higher partial waves.
The measurements were done
at a proton energy of 2.85 GeV, i.e., 83 MeV above the \f\
production threshold, where it was found that
\begin{equation}
R_{pp}= \frac{\sigma(p p \to pp\phi)}{\sigma(p p \to pp\omega)} =
   (3.7 \pm 0.7 ^{+1.2}_{-0.9})\cdot10^{-3} \label{pp}
\end{equation}
In this region, the
correction due to the difference in $\phi pp$ and $\omega pp$ phase
spaces is quite high, and
the ratio (\ref{pp}) becomes a factor 10 larger then the OZI--rule
prediction when corrected.
Therefore, a substantial OZI violation has been observed
also in the proton-proton interaction, and
extracting the ratio from the S-wave only would
lead to an even larger value of the $\phi/\omega$ ratio.
Since the S-wave spin-triplet initial state
becomes diluted at higher energies above the
threshold, one would expect that the deviation
from the \na\ OZI rule should increase
further near threshold.
\footnote{Although the difference in the production
mechanisms of the \f\ and \om\ mesons is obvious, it may not be a
unique signal for intrinsic strangeness, since \f\ meson production near
threshold is anyway expected to be in the S-wave state.
It is unclear whether 83 MeV above
threshold is still the region of S-wave dominance.
Experimental measurements of \f\ and \om\ production
nearer threshold are badly needed.}\\
~\\
\noindent
$\bullet$ {\bf Enhancement of \ten\ production in spin--triplet
annihilation states}\\
~\\
\noindent
Recent measurements of
the $\bar{p}p \to K^+K^-\pi^0$ channel at three hydrogen
densities~\cite{Pra.98} provide the possibility
of comparing the yields of the
\ten\ with that of the $f_2(1270)$ meson, which consists of
light quarks only, in both the S and P waves. It turns out
that
\begin{eqnarray}
 R(f_2'(1525)\pi^0/f_2(1270)\pi^0)
 & = &(47 \pm 14 ) \cdot 10^{-3} ~~~~~\mbox{(S-wave)}  \label{f5} \\
 & = &(149 \pm 20 ) \cdot 10^{-3} ~~~\mbox{(P-wave)}  \label{f6}
\end{eqnarray}
By comparison, on the basis of the \na\
OZI rule one would expect that the ratio
should be of the order of $R(f'_2 /f_2)=(3-16)\cdot10^{-3}$.
The S-wave result (\ref{f5}) is consistent with the Crystal Barrel
measurement~\cite{Ams.98}
$R(f_2'(1525)\pi^0/f_2(1270)\pi^0)= (26\pm10) \cdot 10^{-3}$ for
annihilation in liquid hydrogen, where the S wave is dominant.
We regard the excess of \ten\ production (\ref{f5}) observed in the
S-wave as marginal within the experimental errors.
However, a strong apparent violation of the OZI rule is seen
(\ref{f6}) for \an\ from the P-wave,
as predicted by the polarized-strangeness
model~\cite{Ell.95}.\\
~\\
\noindent
$\bullet$ {\bf Enhancement of \s\ pseudoscalars from spin--singlet
states}\\
~\\
\noindent
The polarized-strangeness model predicts that the
formation of an \s\ system with $J^{PC}=0^{-+}$ should be
enhanced from the spin--singlet states. However, some
caution is needed when connecting this statement with the
data on the production
of real pseudoscalar mesons, due to their large
departure from ideal mixing.

The OBELIX collaboration \cite{Nom.98} recently measured
$\bar pp$ annihilation at rest into the $\phi\eta$
final state for
liquid hydrogen, and for gas at NTP and at a low pressure  of $5$ mbar.
The $\phi\eta$ final state has the same $J^{PC}$ as the
$\phi\pi^0$ final state, so one might have expected to see the same
selection rule as (\ref{3s})-(\ref{1p}).
However, the opposite trend is actually seen:
the yield of the $\bar{p}p \to \phi \eta$ channel grows
with decreasing target density, and
the corresponding branching ratios are
\begin{eqnarray}
B.R.(\bar{p}p \to\phi\eta, ^3S_1) & = &(0.76\pm 0.31)\cdot10^{-4} \label{r1} \\
B.R.(\bar{p}p\to\phi\eta, ^1P_1) & = &(7.72\pm 1.65)\cdot10^{-4}
\label{r2},
\end{eqnarray}
another challenge for model interpretations.

Prediction of the rate of $\phi \eta$ production
is not straightforward in the
framework of the polarized intrinsic-strangeness model.
Since the $\eta$ meson has a substantial $\bar{s}s$ component, there
is a contribution to the
production of the $\phi \eta$ final state via
the production of two $\bar{s}s$ pairs, one in the spin-triplet state
and the other in the spin-singlet state.
If one treats the reaction $\bar{p}p \to \phi\eta$ as the formation
of {\it pseudoscalar} \s\ strangeonium, then the polarized
intrinsic-strangeness model predicts that it should be formed from the
{\it spin-singlet} initial state. It would be interesting  to measure
the density dependence
of the $\bar{p} p  \to \omega \eta$ channel. So far,
this yield was measured only
for annihilation in liquid~\cite{Ams.98}, and is quite high:
$Y(\omega \eta) = (1.51\pm0.12)\cdot10^{-2}$. If the arguments
of the polarized strangeness model are valid for $\eta$ production, then
the yield of the
$Y(\omega \eta)$ final state from the $^1P_1$ initial state should be
higher than from the $^3S_1$ initial state.
However, no significant apparent OZI--violation for
the {\it vector} meson ratio $\phi\eta/\omega\eta$ is expected to occur
even for \an\ from the $^1P_1$ state.

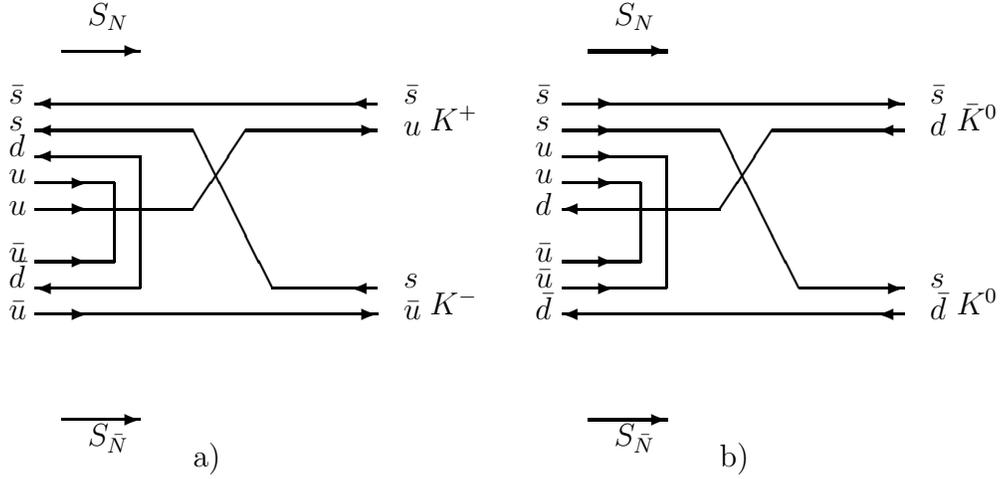
\begin{figure}[htb]

% Picture of  ss shake-out for negative polarization

\linethickness{0.5mm}
\setlength {\unitlength} {0.7mm} \thicklines
\begin{picture}(180,70)(0,10)

%proton 1, low

\put(10,30){\vector(1,0){10}}
\put(20,30){\line(1,0){50}}
\put(15,35){\vector(-1,0){5}}
\put(15,35){\line(1,0){15}}
\put(10,40){\vector(1,0){10}}
\put(20,40){\line(1,0){5}}
%proton 2, upper
\put(70,70){\vector(-1,0){60}}
\put(40,65){\vector(-1,0){30}}
\put(15,60){\vector(-1,0){5}}
\put(15,60){\line(1,0){15}}
\put(10,55){\vector(1,0){10}}
\put(20,55){\line(1,0){5}}
\put(10,50){\vector(1,0){10}}
\put(20,50){\line(1,0){20}}
%annihilation, vertical lines
\put(30,60){\line(0,-1){25}}
\put(25,55){\line(0,-1){15}}
% inclined lines
\put(40,65){\line(1,-2){15}}
\put(40,50){\line(2,3){10}}
%final lines
\put(75,70){\vector(-1,0){5}}
\put(70,65){\vector(1,0){5}}
\put(70,65){\line(-1,0){20}}
\put(60,65){\line(1,0){10}}
\put(75,35){\vector(-1,0){5}}
\put(70,35){\line(-1,0){15}}
\put(70,30){\vector(1,0){5}}
\put(60,30){\line(1,0){15}}
%texts
\put(40,1){a)}
\put(5,70){$\bar{s}$}
\put(5,65){$s$}
\put(5,60){$d$}
\put(5,55){$u$}
\put(5,49){$u$}
\put(5,40){$\bar{u}$}
\put(5,35){$\bar{d}$}
\put(5,29){$\bar{u}$}

\put(80,70){$\bar{s}$}
\put(80,64){$u$}
\put(80,29){$\bar{u}$}
\put(80,35){$s$}
\put(85,65){$K^{+}$}
\put(85,30){$K^{-}$}
\put(20,85){$S_{N}$}
\put(20,5){$S_{\bar{N}}$}
%N1 and N2 spins

\put(15,10){\vector(1,0){15}}
\put(15,80){\vector(1,0){15}}

%\end{picture}
% Picture of  ss shake-out for positive polarization

%\setlength {\unitlength} {0.7mm} \thicklines
%\begin{picture}(80,70)(0,10)

\put(115,30){\vector(-1,0){5}}
\put(115,30){\line(1,0){60}}

\put(110,35){\vector(1,0){10}}
\put(120,35){\line(1,0){10}}

\put(110,40){\vector(1,0){10}}
\put(120,40){\line(1,0){5}}

%proton 2, upper
\put(110,70){\vector(1,0){10}}
\put(120,70){\line(1,0){50}}

\put(110,65){\vector(1,0){10}}
\put(120,65){\line(1,0){20}}

\put(110,60){\vector(1,0){10}}
\put(120,60){\line(1,0){10}}

\put(110,55){\vector(1,0){10}}
\put(120,55){\line(1,0){5}}

\put(115,50){\vector(-1,0){5}}
\put(115,50){\line(1,0){25}}

%annihilation, vertical lines
\put(130,60){\line(0,-1){25}}
\put(125,55){\line(0,-1){15}}
% inclined lines
\put(140,65){\line(1,-2){15}}
\put(140,50){\line(2,3){10}}
%final lines
\put(170,70){\vector(1,0){5}}

\put(175,65){\vector(-1,0){5}}

\put(170,65){\line(-1,0){20}}
\put(160,65){\line(1,0){10}}

\put(170,35){\vector(1,0){5}}
\put(170,35){\line(-1,0){15}}
\put(175,30){\vector(-1,0){5}}
\put(170,30){\line(-1,0){15}}
%texts
\put(140,1){b)}
\put(105,70){$\bar{s}$}
\put(105,65){$s$}
\put(105,60){$u$}
\put(105,55){$u$}
\put(105,49){$d$}
\put(105,40){$\bar{u}$}
\put(105,35){$\bar{u}$}
\put(105,29){$\bar{d}$}

\put(180,70){$\bar{s}$}
\put(180,64){$d$}
\put(180,29){$\bar{d}$}
\put(180,35){$s$}
\put(185,65){$\bar{K}^0$}
\put(185,30){$K^{0}$}
\put(120,85){$S_{N}$}
\put(120,5){$S_{\bar{N}}$}
%N1 and N2 spins
\linethickness{0.5mm}
\put(115,10){\vector(1,0){15}}
\put(115,80){\vector(1,0){15}}
%\end{picture}

\end{picture}
\vspace*{1cm}
\caption{\it Production of $K\bar{K}$ due to shake-out  of a polarized
\s\ pair in the proton wave function
in $\bar{p}p$ annihilation from an initial $^3S_1$ state,
for (a) negative and (b) positive polarization of the \s\ pair.
The arrows show the directions of the spins of the nucleons and the
quarks.}
\label{kns}
\end{figure}

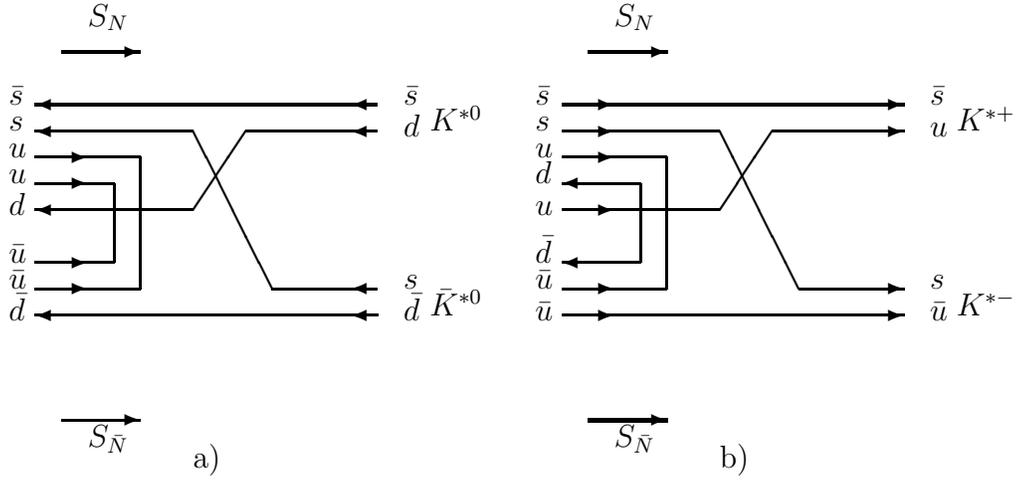
\begin{figure}[htb]

% Picture of  ss shake-out for negative polarization

\linethickness{0.5mm}
\setlength {\unitlength} {0.7mm} \thicklines
\begin{picture}(180,70)(0,10)

%proton 1, low

\put(15,30){\vector(-1,0){5}}
\put(15,30){\line(1,0){60}}
\put(10,35){\vector(1,0){10}}
\put(20,35){\line(1,0){10}}
\put(10,40){\vector(1,0){10}}
\put(20,40){\line(1,0){5}}
%proton 2, upper
\put(70,70){\vector(-1,0){60}}
\put(40,65){\vector(-1,0){30}}
\put(10,60){\vector(1,0){10}}
\put(20,60){\line(1,0){10}}
\put(10,55){\vector(1,0){10}}
\put(20,55){\line(1,0){5}}
\put(20,50){\vector(-1,0){10}}
\put(20,50){\line(1,0){20}}
%annihilation, vertical lines
\put(30,60){\line(0,-1){25}}
\put(25,55){\line(0,-1){15}}
% inclined lines
\put(40,65){\line(1,-2){15}}
\put(40,50){\line(2,3){10}}
%final lines
\put(75,70){\vector(-1,0){5}}
\put(75,65){\vector(-1,0){5}}
\put(70,65){\line(-1,0){20}}
\put(60,65){\line(1,0){10}}
\put(75,35){\vector(-1,0){5}}
\put(70,35){\line(-1,0){15}}
\put(75,30){\vector(-1,0){5}}
\put(70,30){\line(-1,0){15}}
%texts
\put(40,1){a)}
\put(5,70){$\bar{s}$}
\put(5,65){$s$}
\put(5,60){$u$}
\put(5,55){$u$}
\put(5,49){$d$}
\put(5,40){$\bar{u}$}
\put(5,35){$\bar{u}$}
\put(5,29){$\bar{d}$}

\put(80,70){$\bar{s}$}
\put(80,64){$d$}
\put(80,29){$\bar{d}$}
\put(80,35){$s$}
\put(85,65){$K^{*0}$}
\put(85,30){$\bar{K}^{*0}$}
\put(20,85){$S_{N}$}
\put(20,5){$S_{\bar{N}}$}
%N1 and N2 spins

\put(15,10){\vector(1,0){15}}
\put(15,80){\vector(1,0){15}}

%\end{picture}
% Picture of  ss shake-out for positive polarization

%\setlength {\unitlength} {0.7mm} \thicklines
%\begin{picture}(80,70)(0,10)

\put(110,30){\vector(1,0){10}}
\put(120,30){\line(1,0){55}}

\put(110,35){\vector(1,0){10}}
\put(120,35){\line(1,0){10}}

\put(120,40){\vector(-1,0){10}}
\put(120,40){\line(1,0){5}}

%proton 2, upper
\put(110,70){\vector(1,0){10}}
\put(120,70){\line(1,0){50}}

\put(110,65){\vector(1,0){10}}
\put(120,65){\line(1,0){20}}

\put(110,60){\vector(1,0){10}}
\put(120,60){\line(1,0){10}}

\put(120,55){\vector(-1,0){10}}
\put(120,55){\line(1,0){5}}

\put(110,50){\vector(1,0){10}}
\put(120,50){\line(1,0){20}}

%annihilation, vertical lines
\put(130,60){\line(0,-1){25}}
\put(125,55){\line(0,-1){15}}
% inclined lines
\put(140,65){\line(1,-2){15}}
\put(140,50){\line(2,3){10}}
%final lines
\put(170,70){\vector(1,0){5}}

\put(170,65){\vector(1,0){5}}

\put(170,65){\line(-1,0){20}}
\put(160,65){\line(1,0){10}}

\put(170,35){\vector(1,0){5}}
\put(170,35){\line(-1,0){15}}
\put(170,30){\vector(1,0){5}}
\put(170,30){\line(-1,0){15}}
%texts
\put(140,1){b)}
\put(105,70){$\bar{s}$}
\put(105,65){$s$}
\put(105,60){$u$}
\put(105,55){$d$}
\put(105,49){$u$}
\put(105,40){$\bar{d}$}
\put(105,35){$\bar{u}$}
\put(105,29){$\bar{u}$}

\put(180,70){$\bar{s}$}
\put(180,64){$u$}
\put(180,29){$\bar{u}$}
\put(180,35){$s$}
\put(185,65){$K^{*+}$}
\put(185,30){$K^{*-}$}
\put(120,85){$S_{N}$}
\put(120,5){$S_{\bar{N}}$}
%N1 and N2 spins
\linethickness{0.5mm}
\put(115,10){\vector(1,0){15}}
\put(115,80){\vector(1,0){15}}
%\end{picture}

\end{picture}
\vspace*{1cm}
\caption{\it Production of $K^*\bar{K}^*$ due to shake-out  of a
polarized \s\ pair from the proton wave function
in $\bar{p}p$ interaction from an initial $^3S_1$ state,
for (a) negative and (b) positive polarization of the \s\ pair.
The arrows show the directions of the spins of the nucleons and the
quarks.}
\label{sh}
\end{figure}

It is interesting that similar
strong enhancements of $\eta$ production from initial spin-singlet
states were observed in the reactions
$pp\to pp\eta$ and $pn\to pn\eta$~\cite{Cal.98}. An attempt
to interpret this effect in the polarized~strangeness model
was made in~\cite{Rek.97}. There it was pointed out that, at threshold,
the ratio between $\eta$ production on neutron
and on proton is:
\begin{equation}
R_{\eta}=\frac{\sigma(np \to np\eta)}{\sigma(pp \to pp\eta)} =
\frac{1}{4} (1+ \frac{|f_0|^2}{|f_1|^2})             \label{reta}
\end{equation}
where $f_1$ and $f_0$  are the amplitudes corresponding to
total isospin $I=1$ and $I=0$, respectively. At threshold,
when the orbital momentum of
two nucleons in the final state is $l_1=0$ and the orbital momentum of the
produced meson relative to the center-of-mass system of these two
nucleons also vanishes, the connection between the isospin and the total
spin of the two nucleons in the initial state is fixed.
The amplitude $f_1$ corresponds to the spin-triplet
initial nucleon state, and
the amplitude $f_0$ corresponds to the spin-singlet one.
Therefore, using the experimental data on the $pp$ and $np$ cross sections,
it is possible to estimate
the ratio between spin-singlet and spin-triplet amplitudes.
Recent measurements of
$\eta$ production in the threshold region~\cite{Cal.98} show that
the ratio is fairly constant, with a value $R_{\eta}
\approx 6.5$. According to (\ref{reta}), this means that,
as was predicted by the polarized-strangeness model,
the spin--singlet amplitude dominates
$ |f_0|^2/|f_1|^2 \approx 25$.\\
~\\
$\bullet$ {\bf Shake-out  of the \s\ pair and open strangeness
production}\\
~\\
\noindent
Rearrangement diagrams like those in Fig.~\ref{fig:8}
lead to the production of \s\ systems, but the shake-out of the
\s\ pair is also possible, via the diagrams
shown in Figs.~\ref{kns},~\ref{sh}.

\vspace{1cm}

\vspace{1cm}

If the spins of the nucleon and antinucleon
are oriented in the same direction, as, e.g., in the $^3S_1$ initial
state, the shake-out of {\it negatively} polarized \s\
will form preferentially the {\it charged} pseudoscalar $K^+ K^-$
mesons from $s$ and $u$ quarks - which have opposite polarization - and
{\it neutral} vector
$K^{*0} \bar{K}^{*0}$ mesons,
from $s$ and $d$ quarks - which have the same polarization. The
corresponding quark diagrams are shown in Fig. 2a and 3a.
On the other hand,
if the \s\ quarks are polarized {\it positively}, i.e., along the
direction of the nucleon spin,
then $s$ and $u$ quarks will have the same polarization and
they will form preferentially the 
{\it neutral} pseudoscalar $K$ mesons and {\it charged} vector
mesons $K^{*}$,
as seen in Figs.~2b and 3b, respectively.

Therefore, shake--out of the
negatively-polarized
\s\ pair from the $^3S_1$ initial state should lead to the enrichment of
charged $K^+K^-$ pairs over $K^0 \bar{K}^0$ ones and
neutral $K^{*0}\bar{K}^{*0}$ over $K^{*+} K^{*-}$.
On the other hand, these effects should be absent
for \an\ from the spin--singlet initial state
$^1S_0$.

It has been known for a long time (see, e.g.,~\cite{Dos.88}) that
the yield of 
$K^+K^-$ pairs is indeed slightly higher than that
of $K^0 \bar{K}^0$ pairs
for \an~ in liquid, where the $^3S_1$ state dominates:
\begin{eqnarray}
Y(K^+K^-)& = (10.8 \pm0.5)\cdot10^{-4} \label{kk1} \\
Y(K^0 \bar{K}^0) & =  (8.3 \pm0.5)\cdot10^{-4} 
\end{eqnarray}
When combined with the large yield from the pure isospin $I=1$ state
reaction $\bar{p}n \to K^- K^0$~\cite{Bet.69}:
\begin{equation}
Y(K^-K^0) = (14.7 \pm2.1)\cdot10^{-4} \label{kk2}   
\end{equation}
these data reveal a striking hierarchy 
between the isospin amplitudes $T_0, T_1$ for $K\bar{K}$ production 
from the isospin I=0 and I=1 states:
\begin{equation}
T_1(K\bar{K})/T_0(K\bar{K}) \simeq 5-10  \label{isr}   
\end{equation}
This dynamical selection rule is well-known~\cite{Rot.93},
and the dominance of the isospin I=1 amplitude is
usually considered to be due to an initial-state interaction due to an
admixture of the pure isospin I=1 $\bar{n}n$ state in the \p~ wave function~\cite{Car.89,Jac.91}.
The polarized intrinsic-strangeness model provides an
alternative explanation
of the facts  (\ref{kk1})-(\ref{kk2}), and demonstrates that this selection
rule may have the same origin as  
the preferential production of vector 
$K^{*0}\bar{K}^{*0}$ mesons 
over $K^{*+} K^{*-}$.

This phenomenon has indeed been observed in bubble-chamber
experiments~\cite{Bar.65,Bal.65}, where it was 
found that \an~ into
two neutral $K^*$ dominates over charged $K^*$ formation.
For instance, according to~\cite{Bar.65},
$Y(\bar{p}p \to K^{*0}\bar{K}^{*0}) = (30 \pm 7)\cdot10^{-4}$, whereas
$Y(\bar{p}p \to K^{*+}\bar{K}^{*-}) = (15 \pm 6)\cdot10^{-4}$.

This tendency has recently been confirmed
by the
Crystal Barrel collaboration~\cite{Abe.97} in measurements of the
channel $\bar{p}p \to K^0_L K^{\pm} \pi^{\mp} \pi^0$
in \an\ at rest. It was found that, for \an\ from the
$^3S_1$ state, the ratio between neutral and charged $K^*$ production
is
\begin{equation}
\frac{K^*(neutral) \bar{K}^*(neutral)}
{K^*(charged) \bar{K}^*(charged)} \approx 3
\end{equation}
We are not aware of other theoretical arguments that explain this
unexpected selection rule. On the other hand, the polarized-strangeness
model provides a natural explanation of
this effect, making essential use of the sign of polarized pair.
In remarkable consistency with this hypothesis,
this effect is absent for \an\
from the $^1S_0$ initial state, again as it should be for
shake--out of a polarized \s\ pair.\\
~\\
\noindent
$\bullet$ {\bf Momentum transfer dependence of the OZI rule violation}\\
~\\
\noindent
It was conjectured~\cite{Ell.95} that the degree of OZI rule violation
might depend on the momentum transfer. An
answer to this question is provided by the data obtained by the OBELIX
collaboration on the reaction
$\bar{p} + p \to \phi + \pi^+ + \pi^-$,
which was compared with the similar $\omega$ production
reaction~\cite{Roz.96}.
The ratio $R = Y(\phi\pi^+\pi^-) / Y(\omega\pi^+\pi^-)$ for
the annihilation of stopped antiprotons in a gaseous and a liquid hydrogen
target
was measured as a function of the invariant mass of the
dipion system.
If one sums over all events, without any selection on the dipion mass,
the ratio R is at the level of $(5-6) \cdot10^{-3}$, i.e.,
in agreement with the simple prediction of the OZI rule. However, at small
dipion masses $300~ MeV < M_{\pi\pi} < 500~ MeV$,
the degree of OZI rule violation
increases to $R =(16-30) \cdot10^{-3}$.
This should be compared with the ratio $R =(106\pm12) \cdot10^{-3}$  for
the
$\phi\pi^0/\omega\pi^0$ channel for \an\ in liquid~\cite{Ams.98}, which
proceeds from the same $^3S_1$
initial state. Thus, it indeed turns out that, in \f\ production from
the
$^3S_1$ initial state, the degree of OZI-rule violation increases
as the mass of the system created in association with the \f\ decreases,
i.e., as the momentum transfer increases.
It would be interesting to perform a systematic investigation of the
extent to which the
degree of the apparent OZI--rule violation depends on the momentum
transfer.\\
~\\
\noindent
$\bullet$ {\bf Predictions for $\Lambda$ production}\\
~\\
\noindent
It is straightforward to extend the polarized-strangeness model to the
formation of $\bar{\Lambda}{\Lambda}$ and $\phi \phi$
systems. The PS 185 experiment~\cite{Tor.98}
observes a remarkable suppression of the spin-singlet fraction $F_s$
in the $\Lambda \bar \Lambda$ final state:
$ F_s = 0.00014\pm0.00735$.
This is in agreement with the polarized-strangeness model expectations,
which were analysed in~\cite{Alb.95}.
If the \s\ quarks are polarized in the initial state,
it is natural to expect that they will keep the total spin $S=1$ in the
final state and the spin-singlet fraction should be quite small.

The PS 185 collaboration has also measured  the
$\bar{p}p \to \Lambda \bar \Lambda$ channel
in annihilation on a polarized target, to
evaluate the target spin depolarization $D_{nn}$. The
polarized-strangeness
model predicts~\cite{Alb.95} that $D_{nn}$ should be negative.

This prediction also holds for $\Lambda$ production in polarized
proton-proton interactions. Remarkably, the recent results of the
DISTO collaboration~\cite{Bal.99} confirm this prediction.
The spin-transfer coefficient $D_{nn}$ has been measured for the exclusive
reaction $\vec{p} p \to \Lambda K^+ p$ of a polarized proton of 3.67 GeV/c
on an unpolarized proton target.
It was found that, for positive momentum fractions for the $\Lambda$, $x_F
>0$, $D_{nn} \approx -0.4 $ is large and negative.
The value of $D_{nn}$ reflects the fraction of the
normal beam polarization transferred
to the hyperon. The negative sign of $D_{nn}$ means that the component
of the
$\Lambda$ polarization that is correlated with the beam spin is oriented
opposite to the beam spin. As discussed in Section 4. it is clear that
such a correlation is
a consequence of the negative polarization of the strange quarks in the
nucleon.\\
~\\
\noindent
$\bullet$ {\bf Predictions for $\bar p p \to \phi \phi$}\\
~\\
\noindent
The JETSET collaboration has seen an unusually high apparent violation
of the OZI rule in the reaction  $ \bar{p} + p \to \phi +
\phi$~\cite{Lov.97,Eva.98}. The measured cross section of
this reaction
turns out to be 2-4 $\mu$b  for momenta of incoming antiprotons from
1.1 to 2.0 GeV/c. This is two orders of magnitude higher than the value
of 10 nb expected from simple application of the OZI rule.
If this apparent OZI violation is due to the presence of polarized
strangeness in the nucleon, then it was predicted~\cite{Ell.95}
that the $\phi \phi$ system should be produced mainly from the initial
spin--triplet state. Recent data from the JETSET
collaboration~\cite{Lov.97} indeed demonstrate that $\phi \phi$
production is dominated by the initial
spin--triplet state with $2^{++}$.
Moreover, a preliminary analysis~\cite{Lov.95} shows that
final states with total spin $S = 2$ for the $\phi \phi$ system are
enhanced. This fact is explained naturally in the
polarized-strangeness model, on the basis of the
same arguments  as spin--triplet dominance of the $\Lambda \bar{\Lambda}$
system created in \p\ \an\ \cite{Alb.95}.

\section{Other Approaches}

As we have seen, up to now there are no experimental facts which could be
used to rule out the intrinsic polarized-strangeness model, and its many
successes give credence to the approach and stimulate further
investigations, which we discuss in the next Section. However, first we
discuss in this Section the extent to which other models could explain the
experimental facts discussed in the previous Section.

The polarized nucleon strangeness is not
the only possible explanation of each of the experimental facts
discussed. For example,
in case of $\phi \phi$ production,
the `simplest' hypothesis explaining $2^{++}$ dominance is the
presence of a
tensor glueball. The absence of the spin--singlet state in the
$\Lambda \bar{\Lambda}$ system could be reproduced in meson-exchange
models~\cite{Hai.92}. Clearly, these explanations do not
correlate the two sets of data, as does the polarized-strangeness
model.

Of more interest, it has been suggested that the anomalously high yield of
the $\bar{p}p \to \phi\pi^0$
channel could be explained~\cite{Loc.94,Lev.94,Gor.96}
by rescattering diagrams with OZI-allowed transitions
in the intermediate
state, e.g.,  $\bar{p}p \to K^*\bar{K} \to \phi\pi^0$.
Calculations are capable~\cite{Loc.94,Lev.94,Gor.96}
of reasonable agreement with the experimental data on
the $\phi\pi$ yield for annihilation from the S-wave.
However, what is not yet explained in this approach is the
strong dependence of the \f\ yield on the spin of the initial state.
In fact, it is unclear, in any conventional approach which
does not assume polarized intrinsic strangeness in the
nucleon, why the
$\phi\pi$ yield for annihilation from the S-wave is so strong, but
is absent from the P-wave.

Moreover, strong cancellations are expected between different hadronic
loop amplitudes~\cite{Lipkin,Gei.91,Gei.97}. Indeed,
it has been argued~\cite{Lipkin}
that just this feature is the explanation of the
approximate validity of the OZI rule. Selection of
only one type of intermediate state, such as $K^*K$ or $\rho\pi$,
spoils
this delicate cancellation in an uncontrolled way.
The importance of using a {\em complete} set of OZI-allowed hadronic loops
for the calculations of different nucleon strangeness characteristics
has recently been demonstrated in~\cite{Gei.97}.

Finally,  data
from the OBELIX collaboration~\cite{Pra.98,Fil.98}
demonstrate that \f\ production is uncorrelated
with the $K^*K$ channel.
The annihilation yield of the $K^*\bar{K}$ final state
in \ap\ \an\ at rest have been
determined~\cite{Pra.98} at different target densities.
It turns out that the $^1P_1$ fraction of the
$K^* \bar{K}$ annihilation yield is not negligible,
as one would expect in two--step rescattering models.
It is comparable with the
$^3S_1$ fraction, and {\it increases} as the target
density decreases.
This dependence is opposite to that of the $\phi\pi$ yield, which
decreases with the target density.

The \an\ of antineutrons in flight also exhibits different patterns
for the $\phi\pi$ and the $K^*\bar{K}$ final states~\cite{Fil.98}.
Whereas the $\phi \pi$ cross section decreases strongly with
increasing antineutron energy, the
$K^*\bar{K}$ cross section remains essentially flat within the
measured energy interval.

Also inexplicable in the rescattering mechanisms
is the copious production of tensor strangeonium from the P-wave.
The production of $f'_2$ in the reaction
$\bar{p}p \to f'_2 \pi^0$
was calculated in~\cite{Lev.95} via final-state interactions of
$K^*K$ and $\rho\pi$. The production yield of $f'_2$ so obtained
is rather small, about $10^{-6}$, which is about two orders of magnitude
less than the values (\ref{f5})-(\ref{f6})
measured by the OBELIX collaboration~\cite{Pra.98}.

An attempt to calculate the $\phi$ yields in $\bar{p}p$
annihilation at rest in a non-relativistic quark model with a $\bar{s}s$
admixture in the nucleon wave function was made in~\cite{Gut.97}.
There it was assumed that the $\phi$ is produced by shake-out of the
nucleon $\bar{s}s$ component, which implied that the quantum numbers of
the \s~ pair in the nucleon had to be $J^{PC}=1^{--}$, as opposed to
our $^3P_0$ proposal.
 The branching ratios calculated
for the $\phi \eta$ channel were
$B.R.(^3S_1\to\phi\eta)  = (1.4-1.8)\cdot10^{-4}$ and
$B.R.(^1P_1\to\phi\eta)  = (0.15-0.2)\cdot10^{-4}$. Therefore, this
particular shake-out mechanism predicts the $\phi\eta$ yield
from the P-wave should be ten times less
than from the S-wave, whereas experiment observed
exactly the opposite, namely that the P-wave branching ratio is ten times
higher than the S-wave one: see (\ref{r1})-(\ref{r2}).

An interesting possibility considered in~\cite{Mar.97} is that the
final-state interaction (FSI) of two kaons could enhance
$\phi$ production.
Indeed, for annihilation at rest, the phase-space volume in a $\bar{K} K
X$ final
state is limited, the two kaons are created with low relative momenta,
and they could, in principle, fuse into $\phi$ due via FSI. However,
this model does not explain why the FSI effects are stronger for
annihilation from the P-wave than from the S-wave.

It was suggested in~\cite{Zou.95} that the suppression of the $\phi\pi$
yield from the $^1P_1$ state might be connected with some peculiarity of
the protonium
atom which leads to an abnormally low probability to populate this
$^1P_1$ level.
However, if this atomic-physics effect existed, one would expect to
observe the same suppression of the $\phi\eta$ channel from
$^1P_1$ state. However, this is contrary to the
previously-mentioned experimental
results of~\cite{Nom.98}, where {\it enhancement} of
the $\phi\eta$ channel was observed in annihilation from the
$^1P_1$ state.

It seems that approaches based on traditional
ideas~\cite{Loc.94,Lev.94,Gut.97}
are unable to reproduce all the features of \f\ production
and related phenomena observed so far.
On the other hand, the polarized intrinsic-strangeness model~\cite{Ell.95}
not only provides a transparent physical explanation
of the spin effects observed in the production of \f\,\ten\, $\eta$ and
$K^*$ mesons, as well as $\Lambda$ baryons,
but predicts some definite tests. Some of these have already been
confirmed, as discussed in the previous Section, and others we
discuss in the following Section.

\section{New Tests and Predictions}

The polarized-strangeness model has successfully passed through a number
of experimental tests, which gives credence to the new predictions
discussed below. We propose
two general classes of tests -
to verify the mechanism of strangeness shake--out, and to search for
the correlations due to the negative polarization of the intrinsic
nucleon strangeness.\\
~\\
\noindent
$\bullet$ {\bf $\bar K K$ production at low energies}\\
~\\
\noindent
One example how to test the strangeness shake-out mechanism is to study
$\bar K K$ production at low energies.
According to our model, the \s~ pair in the nucleon has preferentially
the quantum numbers
$J^{PC}=0^{++}$. The shake-out of the pair from the nucleon therefore
results in the formation of a $K\bar{K}$
pair with $J^{PC}=0^{++}$ quantum
numbers, which should be preserved by final-state interactions.
This conjecture can be tested, not only in a partial-wave
analysis of $\bar{K}K$ system, but also directly by comparing the relative
yields of
$K_S K_L$ and $K^+ K^-$ pairs created in the hadronic interaction. The
negative C-parity of the $K_S K_L$ system prevents its formation via
shake-out processes: it should be created in the less probable
reaction of rearrangement.

To illustrate this point, let us consider \p\ \an\
at rest into the $\bar{K}K\pi$ channel. One may expect that
\begin{equation}
Y(K^+K^- \pi^0) = 2\cdot Y(K_SK_L \pi^0),
\label{KSKL}
\end{equation}
where we denote by $Y$ the yields of the corresponding channels.
For annihilation in liquid hydrogen, it was found~\cite{Pra.98} that
$Y(K^+K^- \pi^0 ) = (23.7\pm1.6)\cdot10^{-4}$, whereas the yield
of the $K_S K_L \pi^0$ system is~\cite{Ams.98}
$Y(K_SK_L \pi^0) = (6.7\pm0.7)\cdot10^{-4}$.
However, these data refer to the annihilation frequencies of
inclusive $K \bar{K} \pi$ channels, where 
also reactions proceeding via
heavy-meson formation, i.e.,  $\bar{p}p \to \pi M 
(M\to K \bar{K}) \to \pi K \bar{K}$
leading to the same combination of strange and
non-strange mesons contribute.
More data on the exclusive non-resonant $K \bar{K} \pi$ channel are
desirable.\\
~\\

As typical examples of experiments to test the correlations due to
polarization of the strange quarks, we propose the following.\\
~\\
\noindent
$\bullet$ {\bf $\phi$ production
in the interactions of polarized protons with
polarized deuterons}\\
~\\
\noindent
It was predicted~\cite{Ell.95} that  $\phi$ production
in polarized-proton interactions with
polarized deuterons:
\begin{equation}
\vec{p}  + \vec{d} \longrightarrow~ ^{3}He + \phi,     \label{hex}
\end{equation}
would be enhanced when the spins of the proton and deuteron are parallel.
Measurements of the $\phi$ (and $\omega$) yields in the
reaction (\ref{hex}) were performed at Saturne II
using an unpolarized beam and target configuration~\cite{Wur.95}, and a
large deviation from the OZI rule prediction was revealed:
\begin{equation}
R(\phi/\omega) = (80\pm3^{+10}_{-4})\cdot10^{-3}
\end{equation}
This result is promising for future studies of the sensitivity
to polarization of the OZI rule violation in this process.
The main physical advantage of studying the reactions (\ref{hex}),
with $^3He$ production, is  that they offer the possibility to
studying OZI rule violation in the high-momentum-transfer region.

It is remarkable that the standard two--step model of $^3He$ production in
reaction (\ref{hex}) predicts a completely
different behaviour. It was calculated~\cite{Kon.95} that, if the vector
mesons are created via the chain
$pp \to d \pi^+$, $\pi^+ n \to \phi p$ and $d p \to^3He$,
then they should be produced
mainly from the antiparallel orientation of the proton and deuteron spins.
The predicted value for the asymmetry
\begin{equation}
A=\frac{ Y(\uparrow \uparrow) -Y(\uparrow \downarrow)}
{Y(\uparrow \uparrow) + Y(\uparrow \downarrow)},
\label{A}
\end{equation}
where Y is the yield of $^{3}He$ for the parallel and anti-parallel
orientations of the spins of protons and deuterons,
is $A=-0.95$ near threshold. The polarized intrinsic-strangeness
model predicts that $A \approx +1$.\\
~\\
\noindent
$\bullet$ {\bf Tests in nucleon-nucleon interactions}\\
~\\
\noindent
It is important to verify if the selection rules found in
the \ap\ \an\ at rest have counterparts for nucleon-nucleon or
electron-nucleon
interactions.
It is planned at the ANKE spectrometer at COSY~\cite{COSY} to
measure \f\ production in polarized proton interactions with a
polarized proton target:
\begin{equation}
\vec{p} + \vec{p} \to p + p + \phi
\end{equation}
If \f\ production
in the nucleon-nucleon interaction is dominated by the spin--triplet
amplitude, as was observed in \ap\ \an\, then \f\ production should be
maximal when the beam and target nucleons have parallel polarizations, and
suppressed when they are antiparallel.\\
~\\
\noindent
\hbox{
$\bullet$ {\bf Spin dependence of the \f\ production by unpolarized
nucleons}}\\
~\\
\noindent
It is possible to verify the spin dependence of the \f\
production amplitude using unpolarized nucleons:
\f\ production in $np$ and $pp$ collisions
at threshold should also follow (\ref{reta}).
If \f\ is not produced from spin--singlet states,
then the ratio of the $np$ and
$pp$ cross sections at threshold is
\begin{equation}
R_{\phi}=\frac{\sigma(np \to np\phi)}{\sigma(pp \to pp\phi)} =
\frac{1}{4} (1+ \frac{|f_0|^2}{|f_1|^2}) \approx \frac{1}{4}  \label{rfi}
\end{equation}
This ratio was recently calculated~\cite{Tit.97} in the framework of
a one-boson exchange model, i.e., without any assumption about the
nucleon's intrinsic strangeness, with the prediction $R_{\phi}=5$.
Therefore, experimental measurements of this ratio near threshold
could discriminate between the predictions of these theoretical models.\\
~\\
\noindent
$\bullet\,$ {\bf Negative $\Lambda$ polarization in target fragmentation
}\\
~\\
\noindent
The polarized-strangeness model
predicts~\cite{Ek.95} that $\Lambda$ hyperons created in the target
fragmentation region of deep-inelastic scattering should have large
negative longitudinal polarization.
It will be possible to verify this prediction soon in the NOMAD (CERN) and
HERMES (DESY) experiments. Also the
COMPASS experiment at CERN expects~\cite{COMPASS} to study
$\Lambda$ polarization with large statistics, namely of the order of
$10^5$ $\Lambda$ decays.

\section{Conclusions}

We have presented in this paper an update on
polarized intrinsic strange\-ness in the nucleon
wave function. We have presented the latest status
of data from LEAR and other experiments bearing, in particular, on 
large apparent violations of the \na\ 
OZI rule. We have refined and developed a model of the \s\
component of the nucleon wave function, and applied it
to the production of various \s\ quarkonium systems,
such as the $\phi$ and $f_2^\prime (1525)$ mesons.
We have shown that this model describes successfully many
aspects of $p \bar p$ annihilation into final states
containing these mesons, as well as other processes
involving the production of these mesons, $K$ and $K^*$ mesons,
and $\Lambda$ baryons. 
We pointed out the natural connection between our physical picture and
the interpretation of the polarized deep-inelastic scattering data.
 We have also contrasted the predictions of
the polarized intrinsic-strangeness model with alternative
interpretations of the LEAR and related data. Finally,
we have enumerated several new predictions and tests of this
model.

It is clearly desirable to put the polarized intrinsic-strangeness model
on a more solid theoretical basis. Two-dimensional QCD has already
been shown to support the idea that hadronic wave functions
contain \s\ pairs~\cite{Ellis:1992wu}, but this model is not
able to cast
significant light on their possible polarization. Some qualitative
non-perturbative calculations have been proposed in four dimensions (see,
e.g.,~\cite{Koc.98}), but these have yet to become very quantitative.
On the experimental front, the LEAR accelerator has now been
closed. Although interesting analyses of data obtained there are
continuing, the experimental emphasis is necessarily shifting
elsewhere. Fortunately, as we have emphasized in the previous section,
there are several interesting experimental tests that can be made with
low-energy proton accelerators and in deep-inelastic scattering
experiments. We are optimistic that they may bring our
understanding of the strange nucleon wave function to a new level.
\vfill\eject

\section*{Acknowledgments}

The research of one of us (M.K.) was supported in part 
by a grant from the United States-Israel
Binational Science Foundation (BSF), Jerusalem, Israel,
and by the Basic Research Foundation administered by the
Israel Academy of Sciences and Humanities. D.K. 
thanks RIKEN, Brookhaven National Laboratory, and the U.S. Department
of Energy for providing the facilities essential for
the completion of this work.

\end{document}